\documentclass[preprint,authoryear,12pt]{elsarticle}
\usepackage[margin=1.0in]{geometry}
\usepackage{epsfig}
\usepackage{verbatim}
\usepackage{amssymb}
\usepackage{graphics}

\journal{Advances in Space Research}

\newcommand{\Teff}{\mbox{$T_{\rm eff}$}~}

\newcommand{\as}{\mbox{$^{\prime\prime}$~}}
\newcommand{\am}{\mbox{$^{\prime}$~}}

\newcommand{\beqa}{\begin{eqnarray}}
\newcommand{\eeqa}{\end{eqnarray}}

\newcommand{\ebv}{\mbox{$E_{B\!-\!V}$}}

\begin{document}

\begin{frontmatter}



\title{The Ultraviolet View of the Magellanic Clouds from GALEX: A First Look at the LMC Source Catalog}


\author[label1]{Raymond Simons\corref{cor1}}
\author[label1]{David Thilker\corref{cor2}}
\author[label1]{Luciana Bianchi}
\ead[url]{http://dolomiti.pha.jhu.edu}
\author[label2]{Ted Wyder}

\address[label1]{Department of Physics and Astronomy, Johns Hopkins
  University, 3400 N. Charles Street, Baltimore, MD 21218, USA}
\address[label2]{Astronomy Department, Caltech, MC 249-17, 1200 E. California Blvd.,
  Pasadena, CA 91125, USA}
\cortext[cor1]{rsimons@pha.jhu.edu}  
\cortext[cor2]{Catalog correspondence to David Thilker, dthilker@pha.jhu.edu}




\begin{abstract}

  The Galaxy Evolution Exporer (GALEX) has performed unprecedented
  imaging surveys of the Magellanic Clouds (MC) and their surrounding
  areas including the Magellanic Bridge (MB) in near-UV (NUV,
  1771-2831\AA) and far-UV (FUV, 1344-1786\AA) bands at 5\as
  resolution.  Substantially more area was covered in the NUV than
  FUV, particularly in the bright central regions, because of the
  GALEX FUV detector failure. The 5$\sigma$ depth of the NUV imaging
  varies between 20.8 and 22.7 (ABmag).  Such imaging provides the
  first sensitive view of the entire content of hot stars in the
  Magellanic System, revealing the presence of young populations even
  in sites with extremely low star-formation rate surface density like
  the MB, owing to high sensitivity of the UV data to hot stars and
  the dark sky at these wavelengths.

  The density of UV sources is quite high in many areas of the LMC and
  SMC. Crowding limits the quality of source detection and photometry
  from the standard mission pipeline processing.  We performed
  custom-photometry of the GALEX data in the MC survey region
  ($<15^{\circ}$ from the LMC, $<10^{\circ}$ from the SMC). After
  merging multiple detections of sources in overlapping images, the
  resulting catalog we have produced for the LMC contains nearly 6
  million unique NUV point sources within 15$^{\circ}$ and is briefly
  presented herein. This paper provides a first look at the GALEX MC
  survey and highlights some of the science investigations that the
  entire catalog and imaging dataset will make possible.

\end{abstract}

\begin{keyword}
Ultraviolet: surveys; Astronomical Data Bases: catalogs; Galaxies: Magellanic Clouds; Ultraviolet: galaxies; 
\end{keyword}

\end{frontmatter}

\parindent=0.5 cm

\section{Introduction}

As the nearest galaxies to the Milky Way, the Magellanic Clouds (MCs)
have always been targets of considerable interest. The low metallicity
(Olsen et al. 2011, Cole et al. 2005) of the Clouds, combined with
their proximity ($D_{LMC}$ = 50 kpc [Pietrzy$\acute{\rm{n}}$ski et
al. 2013], $D_{SMC}$ = 61 kpc [Hilditch et al. 2005]), allows for
detailed observations and stellar evolution studies probing conditions
rather different than in the Milky Way. However, comprehensive surveys
become demanding and can only be efficiently performed with wide-field
instruments. Large surveys of the Magellanic Clouds have been
accomplished in optical and infrared bands, but a complete catalog of
point sources in the ultraviolet (UV) has long remained missing due to
the difficulty of obtaining the required observations from space.

The Magellanic Clouds have been previously observed in the UV by
imaging and spectroscopic instruments ranging from rocket-borne
cameras to HST, with a great range in resolution and areal coverage.
Page \& Carruthers (1981) first observed the LMC in the wavelength
range $1050-1600\AA$ during the Apollo 16 mission, but at very low
(3\am) resolution. These Apollo observations delineated the overall UV
morphology of the LMC very well and motivated continued efforts. This
pathfinder study was followed by Smith, Cornett, \& Hill (1987) who
imaged the entirety of the LMC in FUV and NUV bands at 50\as
resolution, still too low to resolve individual stars.  Until GALEX,
this work remained the highest quality panoramic imaging for either of
the MCs.  UIT, flown on the Space Shuttle during the Astro-1 and
Astro-2 missions, also targeted the Clouds.  These observations (LMC:
Parker et al. 1998, 2001; SMC: Cornett et al. 1994, 1997; selected
regions: Cheng et al. 1992, Hill et al. 1995) were of much higher
resolution (3\as), even surpassing GALEX data in that respect.
However, only a limited number of fields could be obtained during the
Astro1+2 missions.  FUSE, IUE and HST have provided numerous
spectroscopic and small field-of-view imaging studies of the MCs
(e.g. Brosch et al. 1999, Pradhan et al. 2011).  At the time GALEX was
launched, there remained a significant window for improvement in
coverage and sensitivity over the best existing wide-field UV imaging
(from UIT). The Science Team was keenly aware of this but had to be
conservative in terms of brightness limits for the GALEX detectors,
hence delaying the bulk of MC observations until the end of the NASA
supported GALEX mission.

The GALEX mission (Martin et al. 2005, Morrissey et al. 2007, Bianchi
2009, 2011) has provided wide-field imaging in two UV bands, far-UV
(FUV, 1344-1786\AA) and near-UV (NUV, 1771-2831\AA), with a field of
view of $\approx$1.2$^{\circ}$ diameter and a resolution of
$\approx$4.2/5.3\as (FUV/NUV).  In this paper, we present a first look
from a comprehensive GALEX survey of the Magellanic Clouds and their
environment, particularly focusing on the LMC UV source catalog while
briefly outlining other planned studies.

\section{Data and Coverage}

For the study of the Magellanic System we include observations from
all GALEX surveys (AIS, MIS, NGS, GI, etc., Martin et
al. 2005)\footnote{AIS: All-Sky Imaging (100 s), MIS: Medium Imaging
  (1,500 s), NGS: Nearby Galaxy (1,500 s), GI: Guest Investigator} for
which the planned field center was either within 15$^{\circ}$ of the
LMC or 10$^{\circ}$ from the SMC. These radial limits encompass a
contiguous area of the sky including both Clouds, the entire
Magellanic Bridge (MB), and some of the Magellanic Stream nearest to
the galaxies (Fig. 1). A 'tile' defines the intended pointing of a
GALEX observation, which may be performed with separate exposures
(visits).  All exposures of the same field are later coadded to
improve S/N. We performed photometry on each visit-level image, rather
than on coadds of repeated visits to the same tile.  This allows us to
track variability of sources if present, for those areas of the sky
observed more than once.  Future work will fold in analysis of coadded
data in order to push our catalogs to lower flux limits in selected
regions.

The survey data have a range of exposure times.  We have separated the
overall dataset into two parts: observations which were part of the
AIS (median exposure time of $\sim$150s, 865 visits typically avoiding
the bright areas of the Clouds, by design) and those which were not
(384[294] visits more centrally concentrated on the LMC[SMC]).  The
distribution of visit NUV exposure times is presented in Fig. 2 for
the LMC and SMC separately, along with a map of the accumulated total
exposure across the survey area.

\begin{figure}[t!]
\label{figure1}
\begin{center}
\includegraphics*[scale=0.8]{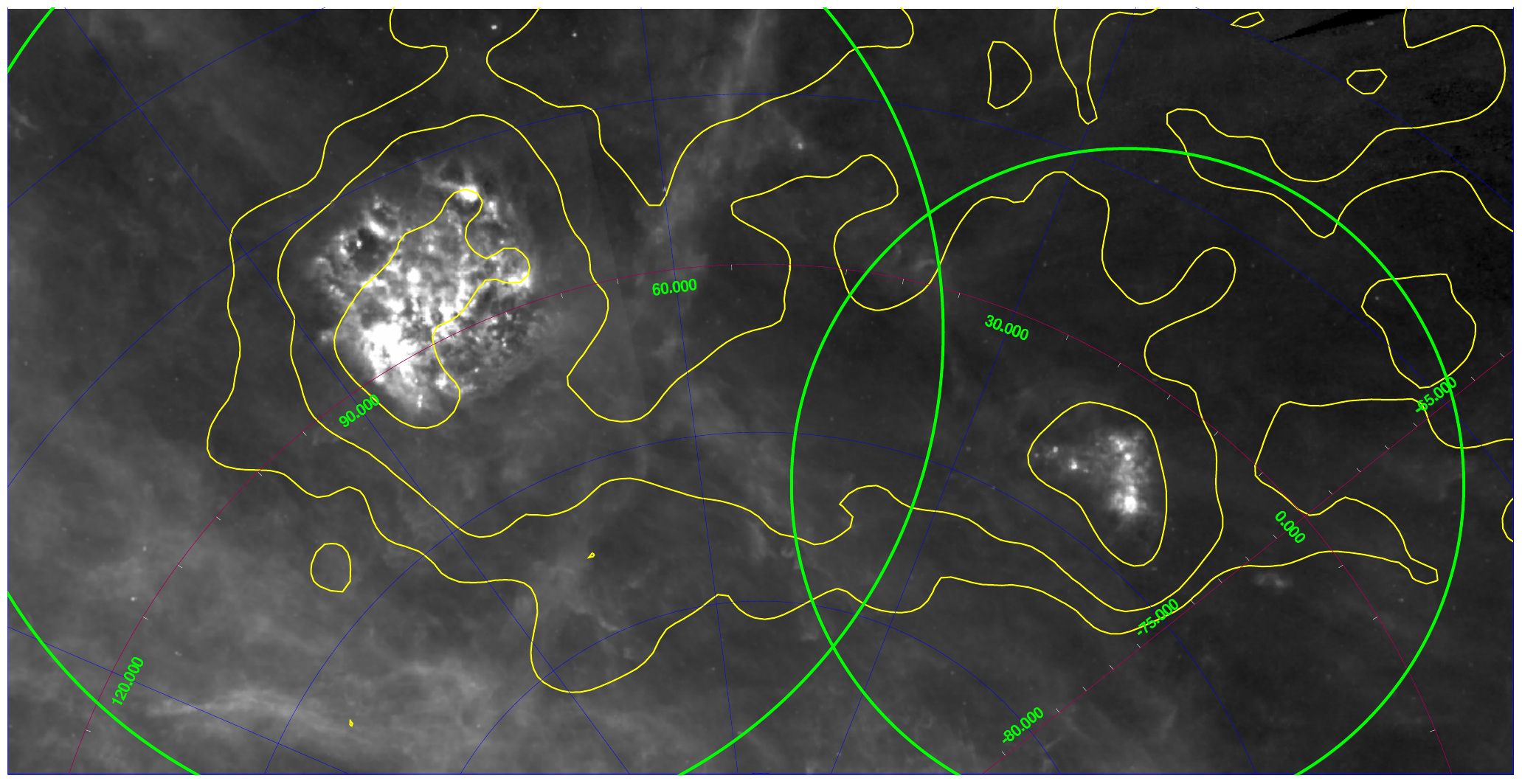}
\end{center}
\caption{The GALEX Magellanic Cloud survey region as seen by IRAS at
  100$\mu m$, contours of the HI distribution of Putman et
  al. (1998).  The N(HI) contours are drawn at 10$^{19}$, 10$^{20}$
  and 10$^{21}$ cm$^{-2}$.  We indicate the radial limits of our
  MC-specific GALEX analysis with the large circles having radius of
  15$^{\circ}$ and 10$^{\circ}$, respectively for the LMC and SMC.
  Note that these limits include the entire Magellanic Bridge. }
\end{figure}

In our catalog result sections (Secs. 4 and 5) , as a preview of
forthcoming products, we only discuss the non-AIS subset pertaining to
the LMC, having a median NUV exposure time of 733s.  This first subset
and source catalog presentation illustrates the type of data and the
reduction (Sec. 3) which will apply to the complete survey.  Full analysis of
the entire Magellanic System catalog (all areas, all exposure depths)
will be published subsequently elsewhere.

The image data were processed with the GALEX pipeline, and taken
directly from the mission visit server at Caltech (internal access for
the Science Team) since they are not yet completely in the MAST (galex.stsci.edu)
public archives.  The data will be identical when they are ingested in
MAST.  Some GALEX imaging presented here includes NUV-only data.
This is because early observations of the MCs frequently encountered
count-rate bright limit safety violations with consequent detector
shutdowns in the FUV, and the FUV detector had already failed by the
time subsequent attempts were acquired with relaxed safety thresholds.
The subset of visits with FUV data are a valuable resource for
characterizing the nature of the sources and inferring their T$_{eff}$
from FUV-NUV color (Section 6.4) even if they are largely confined to
the periphery of the Clouds and MB.

\section{Post-pipeline GALEX Photometry}

In this section we describe the custom photometry processing steps
applied to our GALEX dataset.  As we will show, the GALEX pipeline
measurements are not well-suited to environments as complex as the
Magellanic Clouds.

\begin{figure}[t!]
\label{figure2}
\begin{center}
\includegraphics*[width=12.cm]{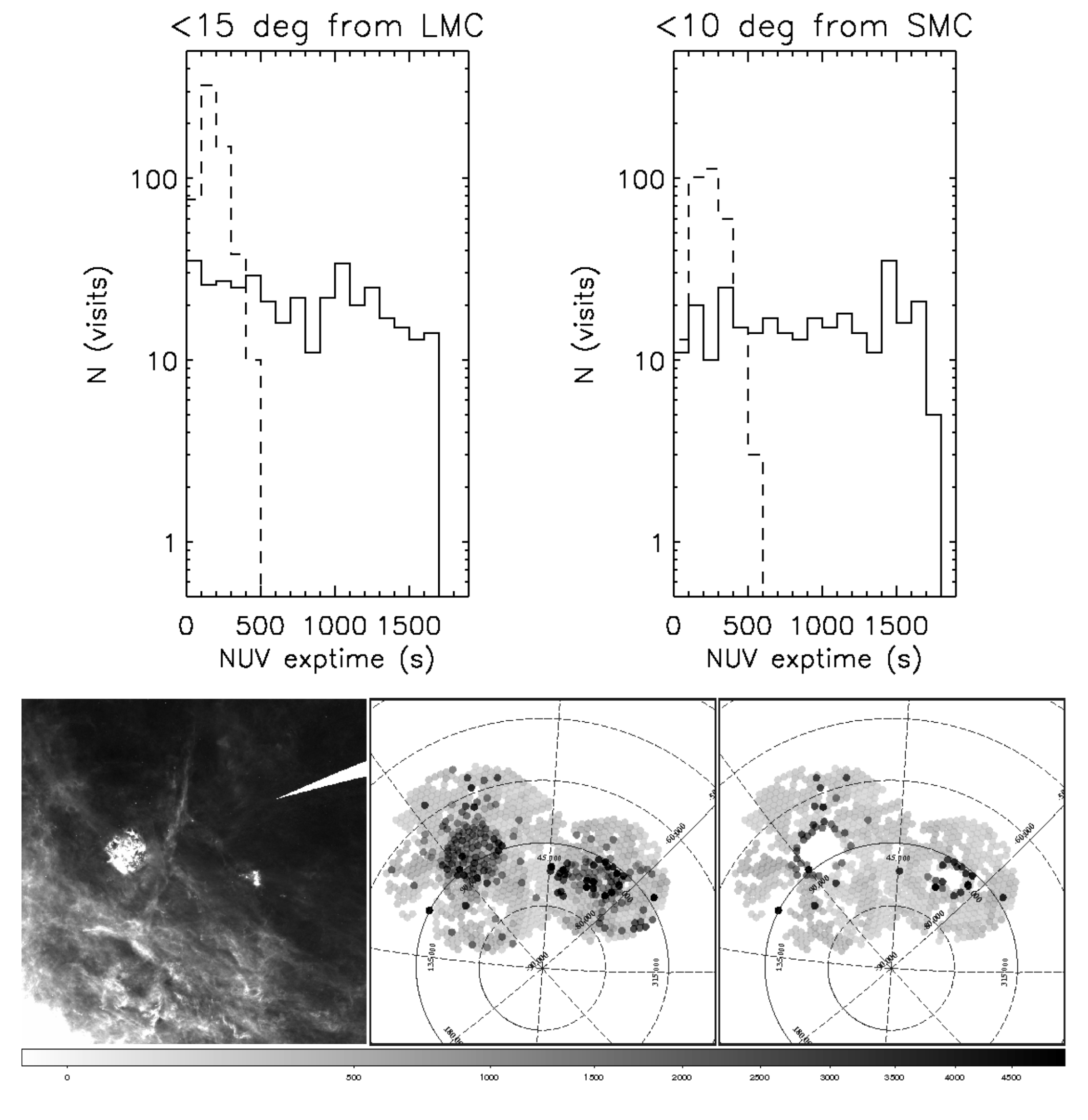}
\end{center}
\caption{Distribution of GALEX exposure times in the NUV-band for
  non-AIS (solid line) and AIS (dashed line) visits falling within
  15$^{\circ}$ of the LMC (top left) or within 10$^{\circ}$ of the SMC
  (top right).  In the image panel, we display the total accumulated exposure time as a function of position for the NUV (bottom center) and FUV (bottom right), in comparison to the IRAS 100$\mu$m survey data (bottom left) over a 55$^{\circ}$-wide field.}
\end{figure}

\subsection{Source Detection}

The GALEX pipeline sometimes fails to resolve closely neighboring
point sources in crowded fields (Figure 3). Often, such partially
blended sources are identified as elongated single sources by the
SExtractor-based pipeline, which was not designed for the level of
crowding encountered in the MCs, even in the peripheral fields.  A
more sophisticated approach had to be used in dense fields with close
and overlapping sources. A NUV-selected source list was generated
using the DAOfind function (implemented in IDL) with a peak amplitude
threshold proportional to the (exptime)$^{-1/2}$ and a NUV-appropriate
Gaussian PSF having FWHM of 5.3\as (3.5 pixels). In practice, the
GALEX PSF varies subtlely from visit to visit and slightly as a
function of position within the field of view (Morrissey et al. 2007).
We further explored this issue, and our approach to dealing with it is
described in the next section.

\subsection{Aperture and PSF Photometry}

Because of the expected variance in the PSF shape discussed above, we
performed both aperture and PSF photometry on the entire set of
detected sources (independently for each visit). Preliminary aperture
photometry on NUV imaging (and FUV when available) was first performed
with apertures of radius 5\as and 10\as.  After applying appropriate
aperture corrections (taken from Morrissey et al. 2007), we found that
the results from both aperture sizes matched well, fully consistent
with most differences expected due to the presence of close neighbors more
frequently falling in the larger aperture.  However, in dense stellar
fields, where crowding is a concern, we need to implement a different
method for performing accurate photometry. In these regions, the need
for PSF photometry becomes critical.  The PSF-fitting method allows us
to dissect significantly blended sources.

The IDL-based PSF-fitting routine we used was developed by T. Small
(personal communication) as a robust deep field method for GALEX
datasets. The source positions adopted for the PSF photometry were
determined on the NUV imaging via DAOfind (as described in
Sec. 3.1). Careful inspection showed the average PSF supplied by the
GALEX project (www.galex.caltech.edu/researcher/techdoc-ch5.html$\#$2)
was not an accurate enough representation of our sources for
individual exposures, often not accounting for slight PSF ellipticity
seen in individual visits, especially for data taken in the late part
of the mission. The variation of the PSF across a single tile (at
least inside a radius of $0.55^\circ$) was not as significant as
intra-visit changes.  For improved photometric accuracy we implemented
an automated method for determining each observation's PSF. To perform
this task, we first ran the PSF fitting code with the GALEX supplied
mission-average PSF. We used the results of this first run to determine sources that
are both bright and isolated enough to represent the PSF. A cut was
made to exclude stars with nearby neighbors, and all other neighbors
within 90\as were masked. The 100 most significant of these stars were
recentered to a reference star. The total flux for each star was
normalized across this subset, maintaining the overall PSF shape. A
new PSF was generated from a datacube of 100 such stars, taking the
median for each pixel. The photometry was then recomputed with the new
empirical PSF. As expected, we find that the custom PSF fitting better
recovers the 5\as aperture photometry than the first run (for isolated
sources), and importantly deblends crowded sources.

\begin{figure}[t!]
\label{figure3}
\begin{center}
\includegraphics*[scale=0.7]{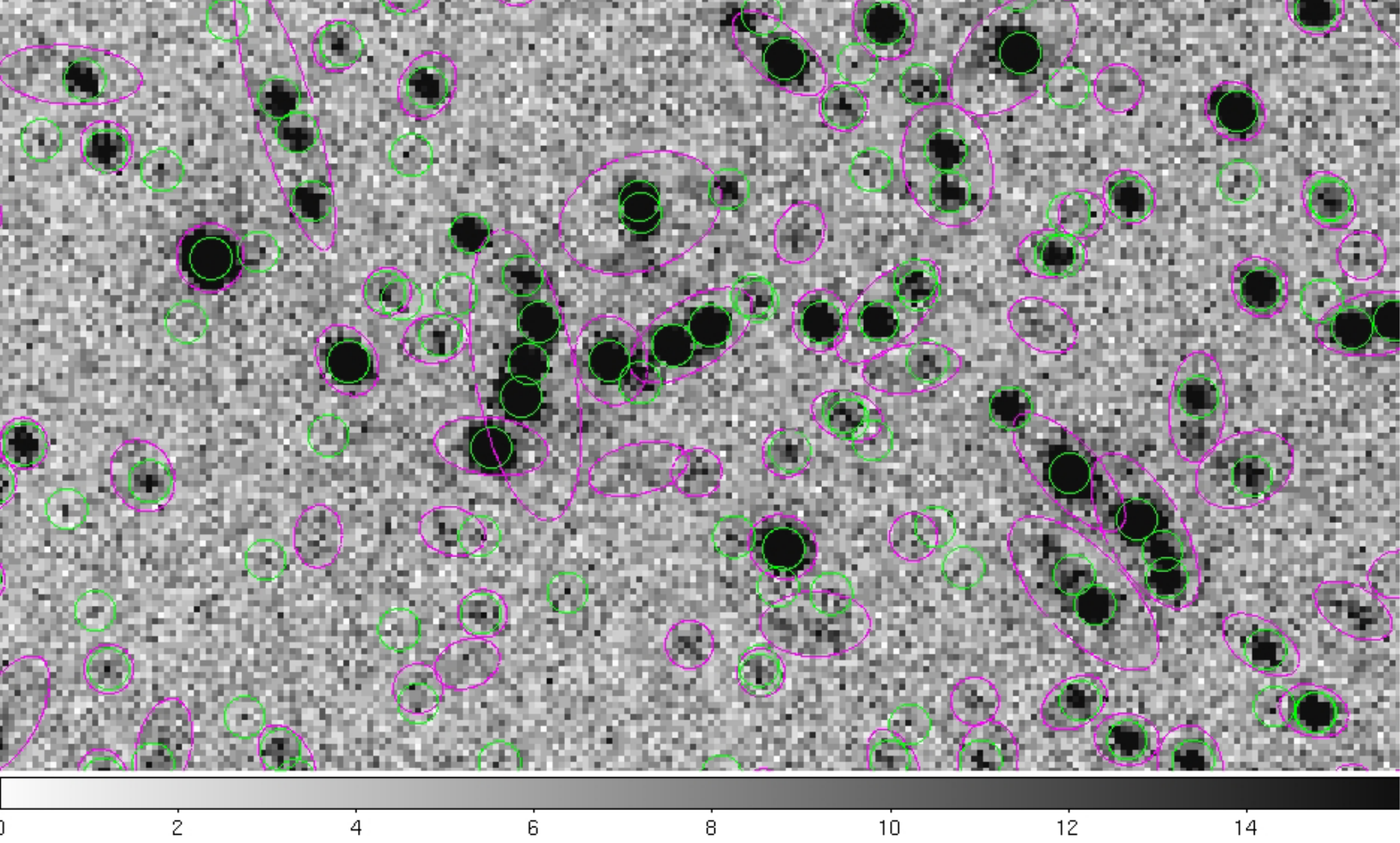}
\end{center}
\caption{Example NUV count rate image with a comparison between
  GALEX-pipeline (purple) and our DAOfind (green) source detections
  for a tile in the LMC (visit exp. 1017s). In crowded fields, the GALEX
  pipeline fails to  separate close and overlapping sources. Crowded
  regions of multiple sources are misidentified as elongated single
  detections by the pipeline.  The DAOfind detections have not yet been
  pruned by significance, sharpness, or roundness at this
  pre-merge (see Sec. 4) stage.}
\end{figure}

\subsection{Photometric Completeness}

The exposure depth and source crowding both vary across our survey,
making it important to estimate completeness limits for separate
regions using artificial star insertion and recovery techniques.  Such
work is now being conducted, but is beyond the scope of this short
contribution.  Standard Galex statistical estimates place the
5$\sigma$ point-source detection limit at NUV $\sim$22.7 and FUV
$\sim$ 22.6 for MIS-depth (1500s) uncrowded observations.  Within the
crowded Magellanic Clouds, we will likely not reach such fiducial
limits for single visit photometry (even for similar exposure times).
For the present catalog, the visit level completeness is estimated
 to be NUV=19.5, FUV=19 AB mag at the shortest exposures
($\sim$100s). However, these short AIS exposures represent a small
fraction of our survey (Fig 2, top).

\section{Description of the Source Catalogs} 

We have constructed comprehensive catalog of unique GALEX sources in
the Large Magellanic Cloud.  A similar product is currently being
generated for the remaining survey area, including all visits
described in Section 2.

After performing photometric measurements for sources detected
independently in each GALEX visit, we had a large database containing
multiple detections of astrophysically unique sources due to the
overlap of adjacent tiles and also stemming from repeated visits.  For
the LMC visits analyzed in depth here (Fig. 4 and Sec. 2), this
concatenated database had nearly 17 million detections.  To produce a
catalog of unique UV sources: (1) we first applied quality cuts on the
concatenated source list, removing detections farther than
0.55$^{\circ}$ from the field center or with DAOfind
roundness/sharpness deviating from the median by greater than
$\pm2\sigma$, together cutting the detection count to 11.3 million,
(2) we then identified sources having repeated measurements within
2.5\as separation, (3) for each set of coincident detections, the one
having the lowest error was retained, except in comparatively rare
cases of an equal error for which the source positioned closest to its
field center was kept, (4) photometry measurements from multiple
detections of each unique source were averaged, (5) finally a
significance cut was applied to retain all unique sources having NUV
error less than 0.5 mag.  The final source catalog contains measured
positions, image quality indicators (sharpness/roundness/crowding),
PSF magnitudes and errors,  plus reference to the individual detections
prior to the catalog merge step.

Our output list of unique UV sources (drawing only from the non-AIS
set of LMC visits) totals 5.8 million.  Although this preliminary set
of analyzed visits is not inclusive of all available data, it does
contain the entire traditional extent of the LMC and can therefore be
considered representative of our final LMC database.  Plans for
public release of this catalog are described in Section 7.

\begin{figure}[t!]
\label{figure4}
\begin{center}
\includegraphics*[width=10cm]{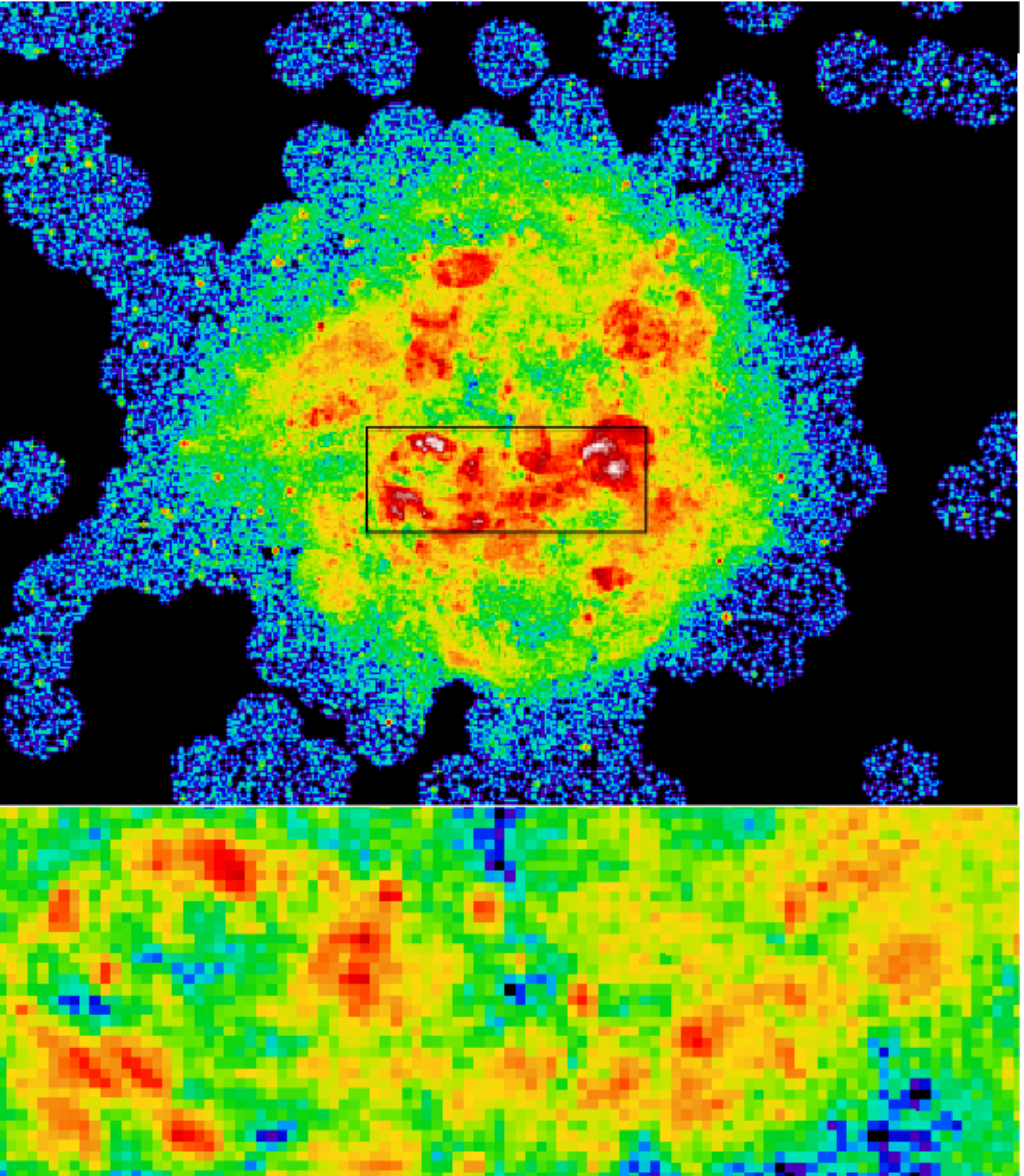}
\end{center}
\caption{Surface density of NUV sources brighter than 19.0 mag in the
  LMC (main panel), evaluated over a $15^{\circ}$ wide field of view
  using only the non-AIS visits analyzed in this contribution.  Some
  arc-like discontinuities are due to different relative completeness
  across tile boundaries with varied exposure time.  The
  inset (boxed) shows a magnified view of a smaller area using a more
  restrictive cut (17.5 mag), helping to delineate small scale
  structures.  Density is estimated across the galaxy counting sources
  in square tesselation bins of 0.04 deg extent per side.  The
  intrinsic stellar density is modulated by local extinction that can
  severely reduce the counts. Levels depicted in these maps are
  discussed in the text.  The detected extent of bright NUV
  point-sources (above the background surface density) is
  approximately equal to the HI extent of the galaxy at N(HI) =
  10$^{20}$ cm$^{-2}$. In this image N is up and E is left, slightly
  different from the orientation in Figs. 1 and 2 (which are on a different projection due to the wider
  area). Comparison is enabled by the coordinate grid shown in
  these earlier plots.}
\end{figure}

\section{The UV Source Content of the LMC Catalog}

Using the list of unique UV sources, we then generated maps of source
surface density to illustrate the overall spatial distribution of
NUV-detected point-like sources in the LMC.  This was accomplished by
counting the number of unique sources brighter than a threshold NUV
magnitude falling into a grid of square sky tassels covering a
15$^{\circ}$-wide field of view centered on the LMC.  We adopted a
tassel size of 0.04 deg per side, corresponding to 35 pc at the
average distance of the LMC.  Work is underway to generate improved
maps incorporating color cuts and an adaptive-size smoothing kernel,
but we present our results for the NUV sources here.  Because the
visits scattered across the galaxy are of varied exposure depth, it is
not straightforward to produce a surface density map without
incompleteness effects becoming apparent if the threshold magnitude is
fainter than the limit associated with the shallowest visits.
Therefore we generated source density maps adopting different choices
of threshold magnitude.  In the lower panel of Fig. 4 we show a
section of the map resulting when a relatively bright threshold of NUV
= 17.5 ABmag is used.  This map does not suffer from discontinuities
due to exposure time differences, and highlights the small scale
structures of star forming complexes.  Such a map could be used to
define star-forming complexes of UV-bright sources (e.g. Kang et
al. 2009, Bianchi et al 2012b, Bianchi et al. 2013, this book). In the
main panel of Fig. 4 we display the surface density map of sources
with NUV $\le$ 19.0 ABmag.  This view illustrates the overall extent
of the LMC as traced by UV-bright sources, but the tile outlines of
the deeper visits begin to become apparent.  In the deeper map, we are
able to trace the galaxy down to a source surface density of 0.09
UV-emitting stars arcmin$^{-2}$ (430 stars kpc$^{-2}$) which yields an
apparent LMC diameter of approximately 10 degrees.  The highest
surface density observed for stars brighter NUV = 19.0 ABmag is
$\sim$50 stars arcmin$^{-2}$ ($2.4\times 10^5$ stars kpc$^{-2}$).  In
some of our deeper visits ($>1000$s), the observed surface density of
{\em all} detected sources approaches levels 10$\times$ greater than
the values we quoted, demonstrating the need for PSF fitting
photometry.  We note that the observed source densities trace the
intrinsic counts but are strongly modulated by extinction, to which UV
fluxes are very sensitive (see also Bianchi 2011 and Bianchi et al.,
this book).  In addition, they represent a lower limit to the actual
number of hot stars, because the crowded cores of star-forming regions
and clusters contain many unresolved sources extended sources that are
excluded from the point-source catalog, and will be treated
seperately.

\section{Future Expansions of the Catalog and Sample Science Applications}

The GALEX UV imaging of the Magellanic System will support a wide
range of science applications, including studies of the point source
population (such as we have already started) but also various
investigations based on unresolved UV features.  We now briefly
describe examples of such studies (e.g. stellar clusters, diffuse
emission).  We further note works which will be made possible by our
future point source catalog expansion (beyond the LMC,
multi-wavelength matching).  Though not discussed here, it will be
instructive to study the multi-wavelength morphology of the LMC and
SMC, contrasted over the UV to IR (Miexner et al. 2006)  and versus
tracers of the gaseous interstellar medium (HI and CO).

\subsection{Stellar Clusters and Star-Forming Complexes}

Much attention has been focused recently on the properties of
young-to-intermediate age star clusters in nearby galaxies.  This
comes from two coupled lines of inquiry: (1) in low mass clusters,
incomplete sampling of the stellar initial mass function at high mass
has implications for the estimation of cluster mass and age (even for
the youngest clusters) due to stochastic fluctuations in predicted/
observed photometric properties and (2) the evolution (dissolution) of
clusters at intermediate ages, hence the viability of dissolved
clusters as a key driver of the field stellar population.  We will
soon undertake analysis of a sample of known young clusters throughout
the Magellanic System, including the Clouds and Bridge.  These objects
are taken from the samples of Baumgardt et al. (2013) for the LMC, and
Bica et al.  (2008) over a wider area.  For each cluster we have
extracted a GALEX postage stamp image and will soon measure the
integrated NUV and FUV (when available) magnitudes.  Together with
existing measurements at longer wavelengths, these will be used to
characterize mass, extinction, and age of each cluster.  In Figure 5,
to illustrate an extreme cluster environment, we show the GALEX
observations of the starburst complex 30 Dor, comparing NUV and
optical imaging in the inset.  In Figure 6 we display a large number
of clusters with more typical masses in the LMC sample (NUV only).  In
Figure 6 it is apparent that the characteristic UV luminosity and
morphology of the clusters varies as a function of both age and mass,
while luminosity must also be modulated by local reddening.

\begin{figure}[h]
\label{figure5}
\begin{center}
\includegraphics*[width=10.cm]{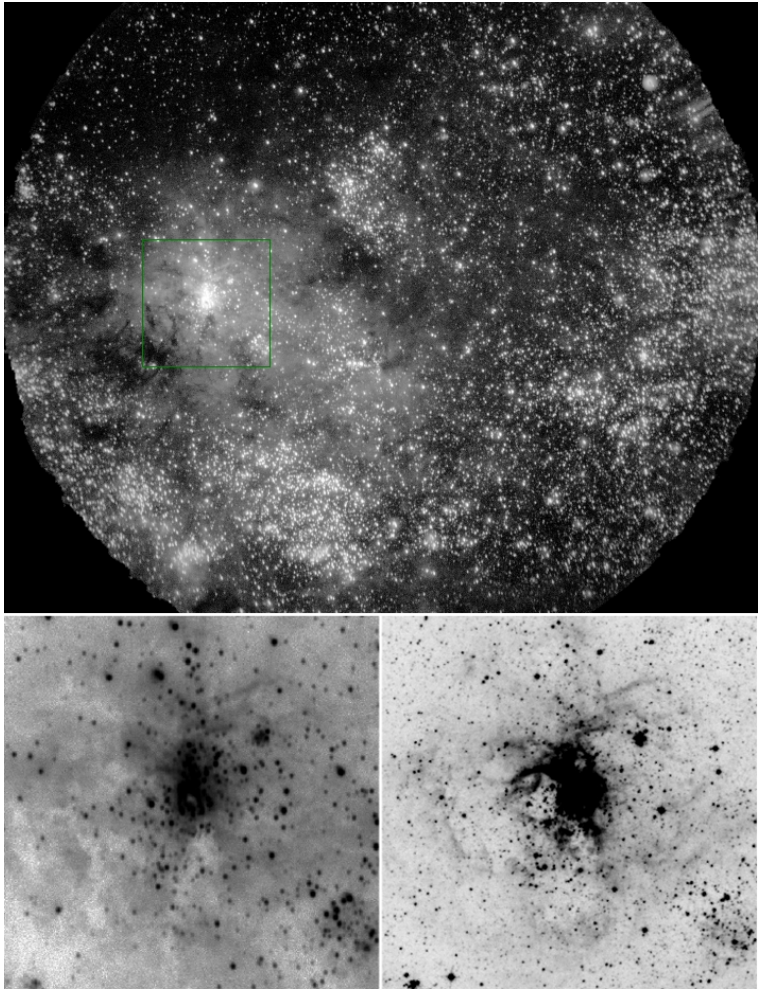}
\end{center}
\caption{One GALEX tile containing 30 Doradus and its environment.  A
  close up view in the UV (left) and optical (right) is provided in
  the inset, and the location of this 0.2$^{\circ}$ (174 pc) box is
  marked in the wide-field image.  The NUV image prominently displays
  the presence of dust, via bright regions of scattered light and also
  in absorption features (especially seen southeast of 30 Dor).}
\end{figure}

\subsection{Diffuse UV Emission}

Inspection of the GALEX imaging for the Magellanic Clouds clearly
reveals the presence of diffuse UV emission filling the space between
resolved stars and clusters.  Figure 5 shows the diffuse
light in the vicinity of 30 Dor as an example.  Much of this
flux is thought to originate as scattered light, correlated roughly
with the amount of dust in the interstellar medium and the UV
continuum from nearby massive stars.  In more distant galaxies similar
diffuse emission is observed, but at the GALEX angular resolution it
is challenging to exclude the possibility of intermediate mass main
sequence stars as a significant contributor to the apparently diffuse
light. The proximity of the MCs offers the opportunity to
accurately decompose the observed UV emission into resolved and
unresolved components, recovering all the point sources producing
appreciable UV.  This is one of the goals of our PSF fitting
photometry.  The ratio of the total flux of the point sources versus
the (sky-subtracted) original image will be used as an indicator of
the observed diffuse fraction, prior to applying an [estimated]
correction for differential extinction between the stars and the
diffuse medium.

\begin{figure}[h]
\label{figure6}
\begin{center}
\includegraphics*[width=10.cm]{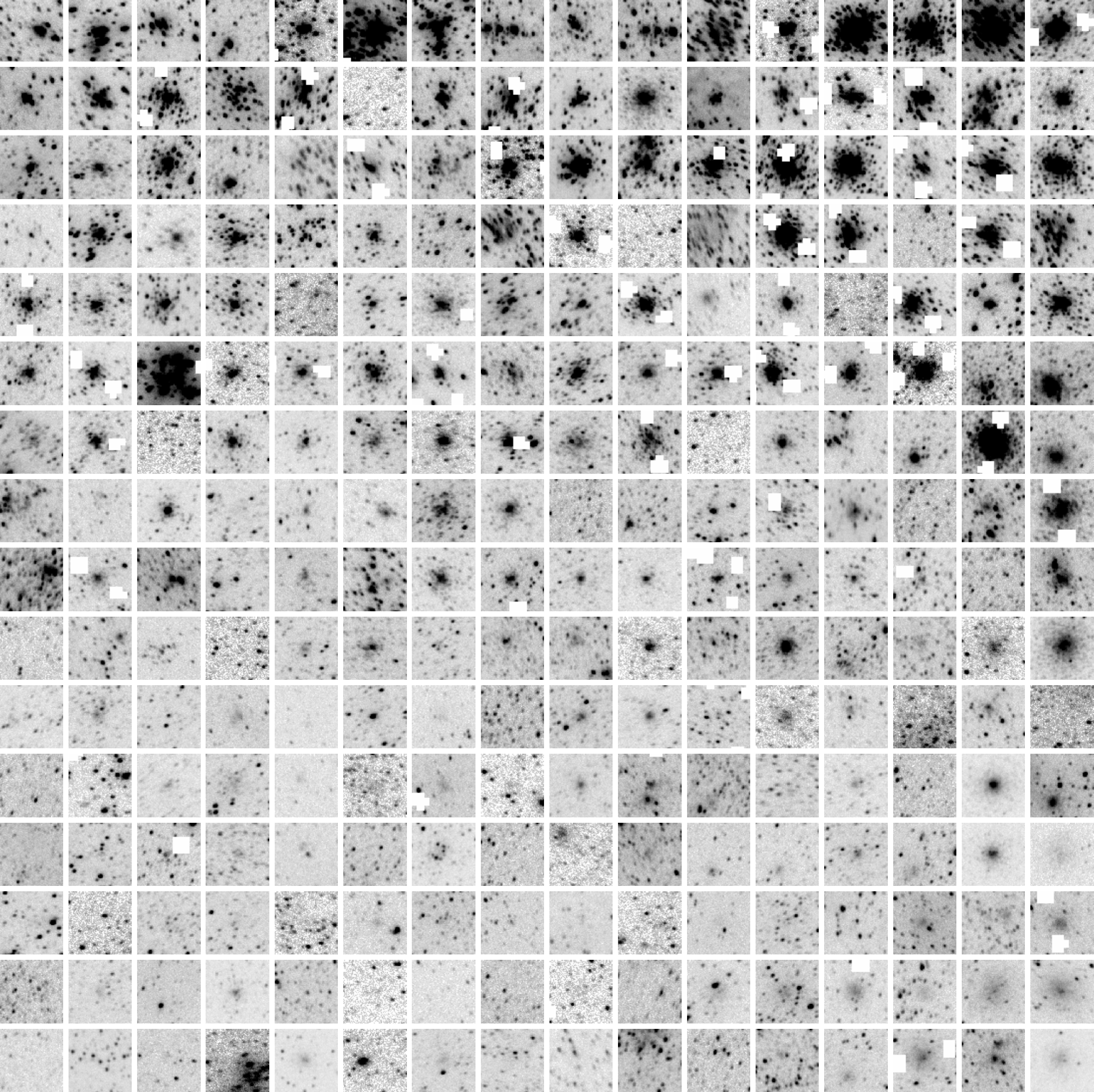}
\end{center}
\caption{NUV images for 256 LMC clusters selected from Baumgardt et
  al. (2013). These are arranged such that the age increases downward
  and (within a row) the mass increases to the right.  The median log
  (age) from Baumgardt et al. corresponding to the rows from top to
  bottom is 7.3, 7.4, 7.6, 7.8, 7.9, 8.0, 8.12, 8.2, 8.3, 8.4, 8.6,
  8.73, 8.97, 9.05, 9.13, and 9.20.  The leftmost column typically
  corresponds to log(M$_{\odot}$) about 3.7 whereas the value for
  rightmost column ranges from 4.3 to 5.2 depending on the observed
  cluster mass function within the age bin.   The field of view shown for each cluster is 38  pc.}
\end{figure}

\subsection{Recent Star Formation in the Magellanic Bridge}

GALEX provides a panoramic view of the MB, enabling a complete census
of its hot stellar population. Our observations (Fig. 7) show that the
Western MB has a significant component of recent star formation
unrepresented by the OB association catalog of Bica \& Schmitt
(1995). We assume contamination by white dwarfs (expected in the
Clouds) is minimal within the MB, because there is no evidence for an
old stellar population from optical studies, even if this is untrue
over our entire survey area.  WD contamination will be explicitly
checked from the UV-optical CMD after matching to ancillary imaging
surveys.  The top panel of Fig. 7 shows an $8^{\circ} \times
3.5^{\circ}$ portion of an FUV mosaic of the Bridge nearest the SMC.
The missing area where the MB joins the SMC has been covered only in
the NUV band.  The importance of the FUV band for UV color selection
of hot sources is apparent in the close-up view of one selected tile,
shown at the bottom of Fig. 7 (with its location circled in the
mosaic).  In a forthcoming paper we will publish a catalog of
UV-selected star-formation complexes and clusters in the MB, utilizing
the type of unique source list described in Section 4 expanded to
cover the entire GALEX MC survey region.

\begin{figure}[h]
\label{figure7}
\begin{center}
\includegraphics* [width=10cm]{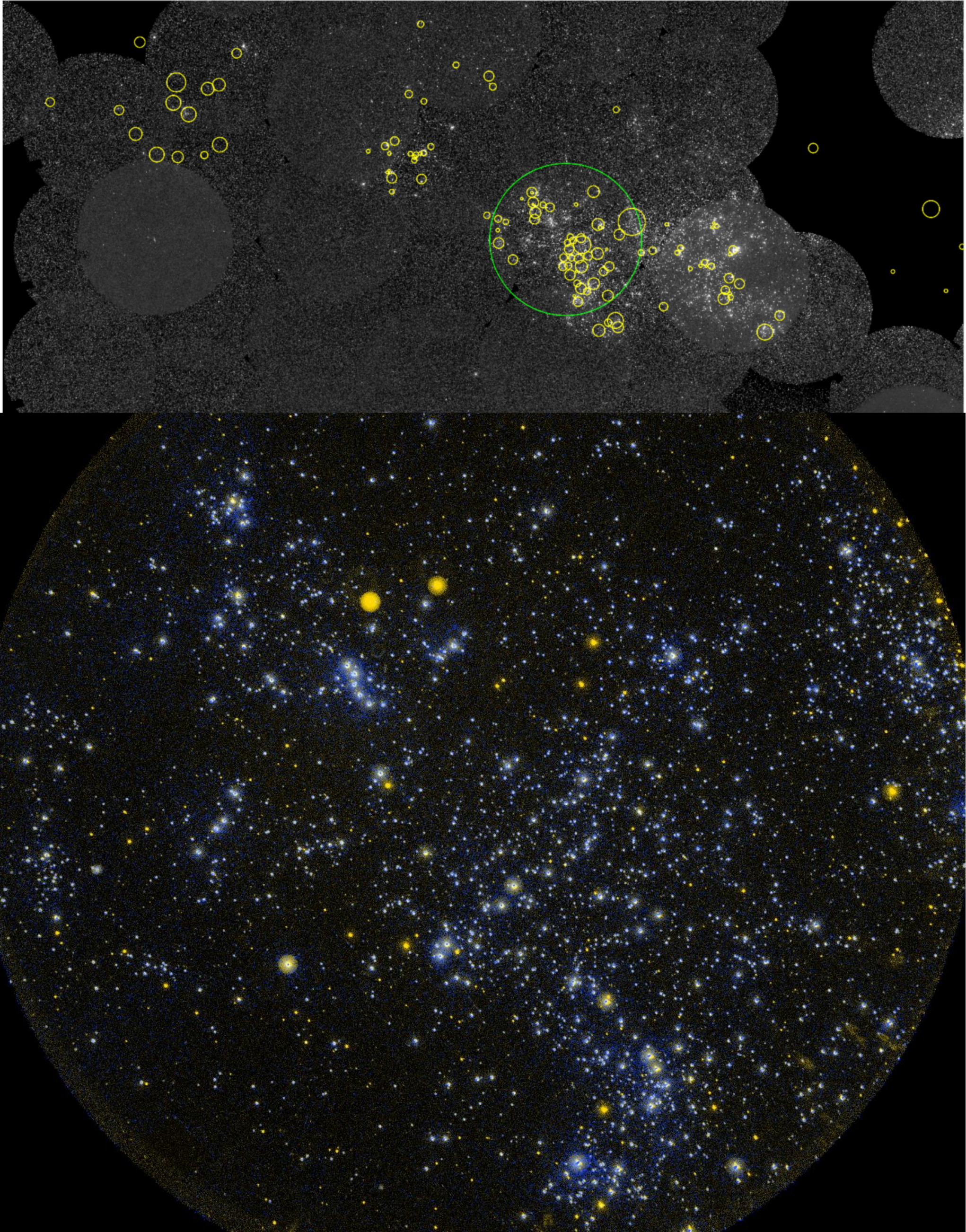}
\end{center}
\caption{A ultraviolet view of the Magellanic Bridge from GALEX.  The top panel shows a mosaic of FUV imaging for several visits and demonstrates the GALEX detection of recent star formation even outside the known OB associations (from Bica \& Schmitt 1995, and marked with yellow circles).  The bottom panel presents a GALEX FUV, NUV (blue/yellow) color composite image for one selected tile.  Its position in the mosaic is marked with a green circle, spanning about 1.1 kpc in diameter.}
\end{figure}

\subsection{Multi-wavelength Catalog Matching and Physical Parameters of the UV sources from UV-optical SED Fitting}

One of the aims of our future work is to estimate the physical
parameters of the detected point-sources by spectral energy
distribution (SED) analysis.  This requires creation of matched
catalogs linking the UV sources to their optical and near-IR
counterparts.  As a pilot study, we matched our UV data to the
 Magellanic Clouds Photometric Survey (Zaritsky et al. 2004),
which covers an 8.5$^{\circ} \times 7.5^{\circ}$ region of
the LMC providing {\it U, B, V,} and {\it I} magnitudes for most stars brighter
than V = 20 Vegamag.  The matching was done online using the CDS
X-Match Service (http://cdsxmatch.u-strasbg.fr/xmatch), adopting a
match radius of 2.5\as.  Because the GALEX images have lower angular
resolution, we tallied instances of multiple matches (more than one
potential optical counterpart per UV source).  UV sources with
multiple optical matches inside the match radius must be treated with
caution, since the UV flux may be the composite of multiple stars.
These are about 20$\%$ of the present catalog.  The GALEX and optical
positions of matched sources are within $\lesssim$0.5\as in the vast
majority ($>90\%$) of cases, indicating consistency in the astrometry
and robust matching between the two catalogs.

GALEX LMC sources with FUV and NUV photometry, and single optical
counterparts (at {\it U B V I}) , were analyzed with grids of stellar model
atmospheres, reddened progressively assuming a variety of extinctions
curves (see Bianchi et al.  2012a, b for details).  The major
parameters derived from SED fitting (through standard $\chi$$^2$
minimization), are the stellar effective temperature \Teff, and the
extinction towards the source, \ebv .  The results depend on the
assumed metallicity, and type of selective extinction
(A$_{\lambda}$/$E_{(B-V)}$); the latter may significantly vary across
different environments. The UV fluxes are particularly sensitive to
this parameter (e.g., Bianchi 2007, 2011), and provide critical diagnostics
for the hottest \Teff's.  Because the distance to the stars is known,
once \Teff and $\ebv$ are derived we can also obtain an estimate of
the radius (and therefore L$_{bol}$) by scaling the best-fit model to
the observed fluxes, accounting for extinction. A few examples of
stellar SED and best-fit models are shown in Figure 8.  We plan to
expand this work to include all regions of both Clouds, even areas
with only NUV coverage.

\begin{figure}[h]
\label{figure8}
\begin{center}
\includegraphics*[width=13.cm]{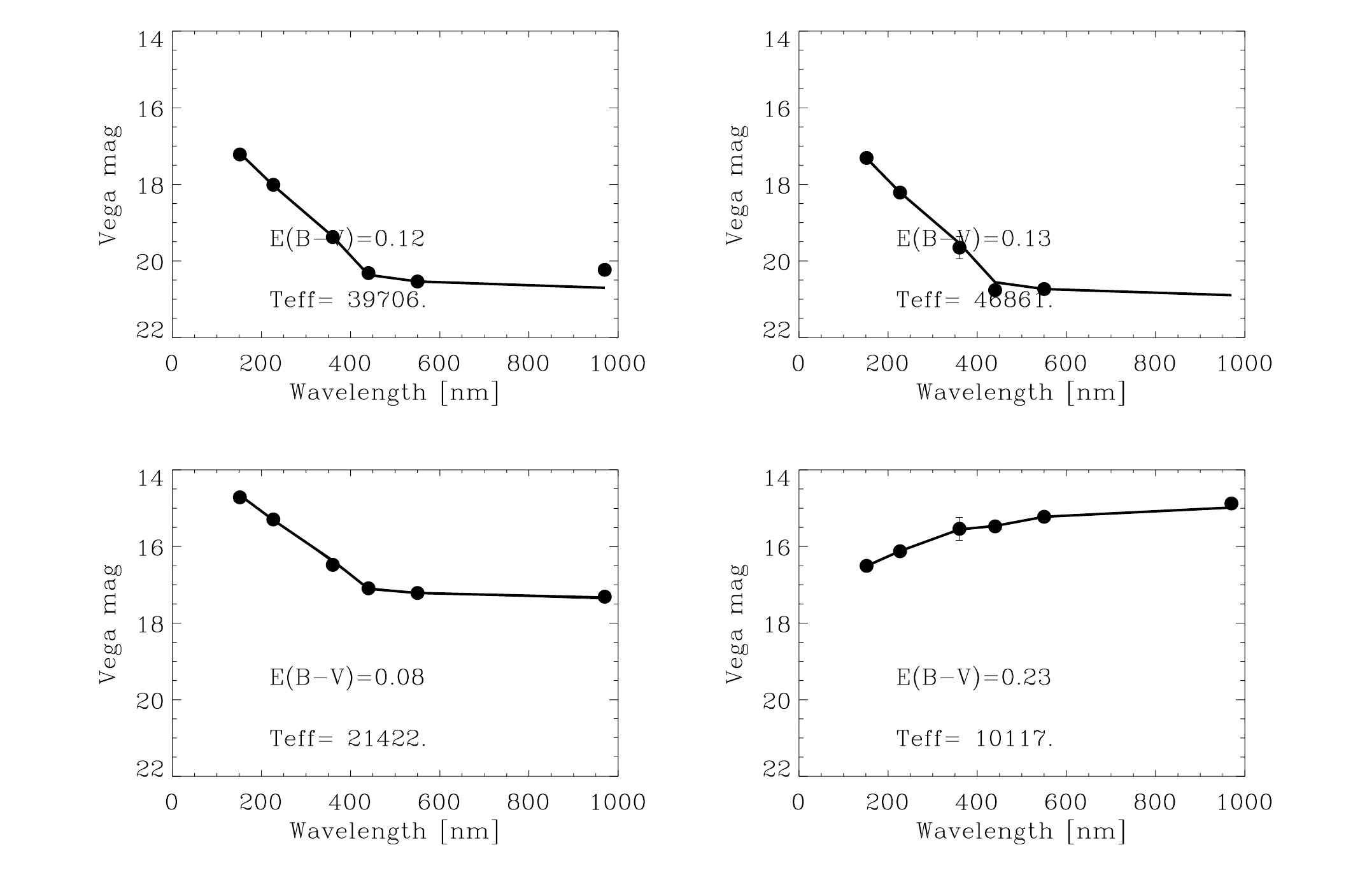}
\end{center}

\caption{UV-optical photometry of stars in the GALEX catalog with
  existing optical data from the MCPS (Zaritsky et al. 2004). Dots are
  the photometric measurements, plotted at the ${\lambda}_{eff}$ of
  the respective bandpasses, the best-fit model magnitudes are
  connected by a line.  Observational errors are smaller than the
  symbol size, with a few exceptions seen in the two righthand
  panels.  The derived \Teff and \ebv  in each case are
  indicated.  Most examples are chosen from the hottest stars.  The
  SED fitting was done using the method of Bianchi et al. (2012a).}
\end{figure}

\section{Summary}

We have constructed a comprehensive catalog of unique GALEX UV sources
in the Large Magellanic Cloud.  A similar product is currently being
generated for the remaining survey area, including all visits
described in Section 2.  The Magellanic Clouds will continue to remain
an active area of interest in studies of low mass galaxy evolution,
morphology and activity. Our survey provides a comprehensive UV
assessment of the Magellanic Clouds, probing active resolved star
formation regions in a low metallicity environment.  The current
contribution was intended to provide a first look at the LMC data, and
highlight important forthcoming efforts based on the complete dataset
(depicted in Figure 2).  Our entire GALEX photometric catalog
(including both LMC and SMC, plus surroundings) will be made available
from the authors' website (dolomiti.pha.jhu.edu) and eventually from
the MAST archive as a high-level science product (HLSP).

~\\
{\bf Acknowledgements: }

We are grateful to the GALEX ``Science Operation and Data Analysis
Team'' for their years of hard work and devotion which led to such a
successful survey mission.  This research has made use of the VizieR
catalogue access tool, CDS, Strasbourg, France.  In particular, the
CDS X-Match service was very helpful.  This research has made use of
the NASA/IPAC Extragalactic Database (NED) which is operated by the
Jet Propulsion Laboratory, California Institute of Technology, under
contract with the National Aeronautics and Space Administration.

~\\

\end{document}